\documentclass[lettersize,journal]{IEEEtran}
\usepackage{amsmath,amsfonts}
\usepackage{cite}
\usepackage{lineno,hyperref}
\usepackage{changes}
\usepackage[ruled, lined, longend, linesnumbered]{algorithm2e}
\usepackage{graphics}
\usepackage{filecontents}
\usepackage{mathtools}
\usepackage{enumitem}
\setlist[itemize]{leftmargin=*}
\usepackage{xcolor}
\begin{document}
	\title{Meta-Learning for Resource Allocation in Uplink Multi-Active STAR-RIS-aided NOMA System}
	\author{Sepideh Javadi, \textit{Graduate Student Member, IEEE}, Armin Farhadi, Mohammad Robat Mili, Eduard Jorswieck, \textit{Fellow, IEEE}, and Naofal Al-Dhahir, \textit{Fellow, IEEE}
		\thanks{Sepideh Javadi and Mohammad Robat Mili are with the Pasargad Institute for Advanced Innovative Solutions (PIAIS), Tehran, Iran (e-mails: sepideh.javadi@piais.ir and mohammad.robatmili@piais.ir). Armin Farhadi is with the School of Electrical and Computer Engineering, College of Engineering, University of Tehran, Tehran, Iran (e-mail: armin.farhadi@ut.ac.ir). Eduard Jorswieck is with the Institute for Communications Technology, Technische Universitat Braunschweig, 38106 Braunschweig, Germany (email: e.jorswieck@tu-bs.de). Naofal Al-Dhahir is with the Department of
		Electrical and Computer Engineering, The University of Texas at Dallas,
		Richardson, TX 75080 USA (aldhahir@utdallas.edu).}}	
	\maketitle
	\makeatletter
	\long\def\@makecaption#1#2{\ifx\@captype\@IEEEtablestring%
		\footnotesize\begin{center}{\normalfont\footnotesize #1}\\
			{\normalfont\footnotesize\scshape #2}\end{center}%
		\@IEEEtablecaptionsepspace
		\else
		\@IEEEfigurecaptionsepspace
		\setbox\@tempboxa\hbox{\normalfont\footnotesize {#1.}~~ #2}%
		\ifdim \wd\@tempboxa >\hsize%
		\setbox\@tempboxa\hbox{\normalfont\footnotesize {#1.}~~ }%
		\parbox[t]{\hsize}{\normalfont\footnotesize \noindent\unhbox\@tempboxa#2}%
		\else
		\hbox to\hsize{\normalfont\footnotesize\hfil\box\@tempboxa\hfil}\fi\fi}
	\makeatother
	\begin{abstract}
	Simultaneously transmitting and reflecting reconfigurable intelligent surface (STAR-RIS) is a novel technology which enables the full-space coverage. In this letter, a multi-active STAR-RIS-aided system using non-orthogonal multiple access in an uplink transmission is considered, where the second-order reflections among multiple active STAR-RISs assist the transmission from the single-antenna users to the multi-antenna base station. Specifically, the total sum rate maximization problem is solved by jointly optimizing the active beamforming, power allocation, transmission and reflection beamforming at the active STAR-RISs, and user-active STAR-RIS assignment. To solve the non-convex optimization problem, a novel deep reinforcement learning algorithm is proposed which integrates Meta-learning and deep deterministic policy gradient (DDPG), denoted by Meta-DDPG. Numerical results reveal that our proposed Meta-DDPG algorithm outperforms the DDPG algorithm with $19\%$ improvement, while second-order reflections among multi-active STAR-RISs provide $74.1\%$ enhancement in the total data rate.    
	\end{abstract}
	\begin{IEEEkeywords}
	Meta-DDPG, NOMA, STAR-RIS, data rate.
\end{IEEEkeywords}
	\addtolength{\abovedisplayskip}{-1.3mm}
	\addtolength{\belowdisplayskip}{-1.3mm}
\section{Introduction}
Reconfigurable intelligent surface (RIS) has emerged as a promising technology for the upcoming sixth generation (6G) wireless systems. However, conventional RIS can only serve wireless devices located on the same side without providing full-space coverage. To address this limitation, the simultaneously transmitting and reflecting (STAR)-RIS has been introduced, enabling simultaneous reconfiguration of both reflected and transmitted signals \cite{53@Mu}.

Non-orthogonal multiple access (NOMA) is another promising solution for enhancing spectral efficiency and achieving massive connectivity by allowing multiple users to share the same resource block \cite{54@Lv}. More specifically, the authors in \cite{SIRS-NOMA} apply NOMA in a multi-carrier RIS-assisted system and aim to maximize the system throughput.
The integration of STAR-RIS with NOMA can enhance the spectrum efficiency and is envisioned as future wireless network structure \cite{1@Zuo, 9@Gao}. For instance, the authors in \cite{1@Zuo} aim to maximize the achievable sum rate in a NOMA-assisted STAR-RIS communication system, where users are classified into various clusters, while the sum rate maximization problem in a STAR-RIS-assisted system, utilizing both cluster-based NOMA and beamformer-based NOMA schemes, is studied in \cite{9@Gao}.

Most of the relevant works have concentrated on enhancing wireless links using one or multiple RISs with a \textit{single signal reflection}. However, few studies address the modeling of multi-RIS deployment scenarios in wireless networks, taking into account \textit{secondary} or \textit{multi reflections} amongst RISs \cite{21@Wang, Single-Multi, secondorder}. 
Particularly, the application of NOMA in multi-RIS downlink system is investigated in \cite{21@Wang}, where the authors aim to minimize the total transmit power. Furthermore, the significance of multi-RIS deployments with multiple signal reflections is studied in \cite{Single-Multi}. In this letter, we consider secondary reflections among multiple active STAR-RISs using NOMA in an uplink transmission to provide enhanced data rate, as the received signal strength can be amplified through second-order reflection \cite{secondorder}. To the best of our knowledge, this is the first work which develops the multi-active STAR-RIS architecture in an uplink transmission with NOMA users, where Meta deep deterministic policy gradient (Meta-DDPG) is applied to solve the problem. 
\section{System Model and Problem Formulation}
As illustrated in Fig. 1, we consider an uplink transmission in a multi-active STAR-RIS-assisted NOMA system comprising of one BS with $N_{\text{BS}}$ antennas and a set $\mathcal{N} = \left\{ {1,2,...,n,...,N} \right\}$ active STAR-RISs, where the $n$-th active STAR-RIS has ${M_n}$ elements denoted by ${\mathcal{M}_n} = \left\{ {1,2,...,{m_n},...,{M_n}} \right\}$. Moreover, the total number of $K$ single antenna users indexed by $\mathcal{K} = \left\{ {1,2,...,k,...,K} \right\}$ are distributed in the system, with each active STAR-RIS serving $K_n$ single antenna users indexed by $\mathcal{K}_n = \left\{ {1,2,...,k_n,...,K_n} \right\}$. 
Multiple active STAR-RISs are distributed within the system, such that the corresponding space of each active STAR-RIS is divided into two parts, namely, reflection space and transmission space. Therefore, the elements of each STAR-RIS divide the incident signal into the reflected signals in the reflection space and the transmission signals in the transmission space. Hereafter, the $k$-th user and the $m$-th element associated with the $n$-th active STAR-RIS are represented by $[k,n]$ and $[m,n]$, respectively.
\begin{figure}
	[ptb]
	\begin{center}
		\includegraphics[width=9.7cm, height=5.5cm]%
		{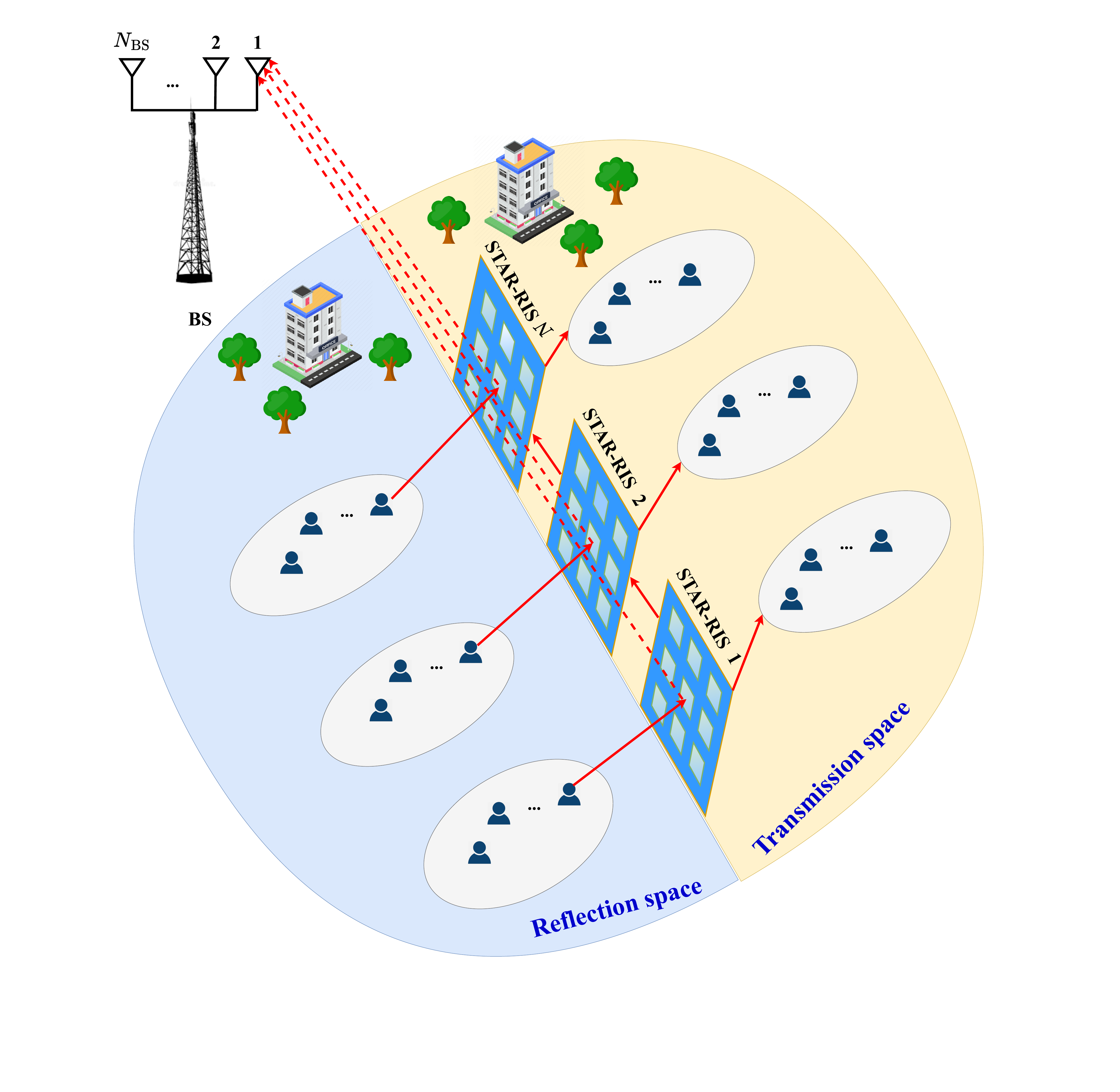}
		\caption{System model.}
		\label{Fig5}
	\end{center}
	\vspace{0cm}
\end{figure}
\subsubsection{Signal Model} 
The incident, reflected, and transmitted signals at the $m$-th element of the $n$-th active STAR-RIS can be expressed as $x_{m,n}$, $r_{m,n} = \sqrt {\delta_{m,n}^r\alpha _{m,n}^r} {e^{j\theta _{m,n}^r}}{x_{m,n}}$, and  $t_{m,n} = \sqrt {\delta_{m,n}^t\alpha _{m,n}^t} {e^{j\theta _{m,n}^t}}{x_{m,n}}$, respectively. Here, the amplification gain, amplitude, and phase shift of the reflection and transmission coefficients for the $m$-th element of the $n$-th active STAR-RIS are denoted by $\left\{ {{\delta_{m,n}^r} ,{\delta_{m,n}^t} \in \left[ {0,\delta_\text{max}} \right]} \right\}$, $\left\{ {{\alpha _{m,n}^r} ,{\alpha _{m,n}^t}  \in \left[ {0,1} \right]} \right\}$ and $\left\{ {\theta _{m,n}^r,\theta _{m,n}^t \in \left[ {0,2\pi } \right)} \right\}$,  respectively. Considering active STAR-RISs, $\delta_\text{max}$ can be greater than 1 to amplify the incident signals \cite{ActiveS}. It is worth noting that contrary to the reflection and transmission phase shifts which are independent, the amplitude coefficients for reflection and transmission are coupled, ensuring that the incident signal energy  equals the sum of the reflected and transmitted signals' energies, i.e., ${\left| {r_{m,n}} \right|^2} + {\left| {t_{m,n}} \right|^2} = {\left| {{x_{m,n}}} \right|^2}, ~\forall m \in \mathcal{M},n \in \mathcal{N}$. Therefore, to realize the above mentioned energy conservation principle, the condition $\alpha _{m,n}^r + \alpha _{m,n}^t = 1$ should be satisfied \cite{56Soleymani}.  
The beamforming vector of the $n$-th active STAR-RIS for the reflection $(l=r)$ and transmission $(l=t)$ spaces is denoted by $\bold{\theta}_n^l = {\left[ {\sqrt {\delta _{1,1}^l\alpha _{1,1}^l} {e^{j\theta _{1,1}^l}},...,\sqrt {\delta_{M,N}^l\alpha _{M,N}^l} {e^{j\theta _{M,N}^l}}} \right]^H}$, where the corresponding diagonal beamforming matrix is defined as $\bold{\Theta}_n^l = \text{diag}\left( {\bold{\theta}_n^l} \right)\in\mathbb{C}^{{M_n} \times {M_n}}$.
Moreover, the association index ${\beta_{k,n}}$ determines if user $k$ is associated with the $n$-th active STAR-RIS which is defined as follows\\

\begin{align}
	\left\{ \begin{array}{l}
		{\beta _{k,n}} = 1,~ \text {if user $k$ is associated with active STAR-RIS $n$},\\
		{\beta _{k,n}} = 0,~ \text {otherwise.}
	\end{array} \right.
\end{align}
We assume that the multi-active STAR-RISs are sorted in descending order based on their distances to the BS, i.e., ${d_{\text{BS},\text{S-RIS}\_1}} > {d_{\text{BS},\text{S-RIS}\_2}} > ... > {d_{\text{BS},\text{S-RIS}\_N}}$.
Hence, the received signal at the BS is given by
\begin{align}
{\bold{y}_{\text{BS}}} = \sum\limits_{n=1}^{ {N}}{\sum\limits_{{k=1}}^{ {K_n}}{\bold{h}_{{k,n}}{\sqrt{{p}_{k,n}}}}} {\hat{x}_{k,n}} + \bold{z},
\end{align}
where $\hat{x}_{k,n}$ represents the transmit signal stemming from the $[k,n]$-th user to the BS with normalized power, i.e., $\mathbb{E}\left\{\left|\hat{x}_{k,n}\right|^{2}\right\}=1$, while ${p}_{k,n}$ denotes the transmit power of the $[k,n]$-th user. Moreover, $\bold{z}$ is the white complex Gaussian noise vector with zero mean and covariance matrix $\sigma^2\bold{I}_{N_{\textrm{BS}}}$, i.e., $\mathbf{z}\sim{\mathcal{C}} \mathcal{N}\left(\bold{0}, \sigma^{2} \bold{I}_{N_{\textrm{BS}}}\right)$, where $\sigma^{2}$ denotes the noise variance and $\mathbf{I}$ is the identity matrix. Furthermore, the diagonal matrix of the $n$-th active STAR-RIS is defined as follows
\begin{align}
\bold{\Theta}_{k,n}=\left\{ \begin{array}{l}
\bold{\Theta} _n^t, \quad \text {if $k$-th user is in the transmission space},\\
\bold{\Theta} _n^r, \quad \text {if $k$-th user is in the reflection space.}
\end{array} \right.
\end{align}
The inter-STAR-RIS reflections are considered, where each active STAR-RIS assists users associated with the farther active STAR-RIS, thus enhancing the transmission quality. In the case of triple or higher-order reflections, where a signal passes through more than two active STAR-RISs, the impact is often negligible due to significant effective path loss. Hence, the equivalent channel coefficients of the $[k,n]$-th user considering second-order reflections for $n=N$ and $n<N$, are, respectively, defined as follows\\
\begin{align}
\begin{array}{l}
\bold{h}_{k,N} = \bold{F}_{N,\text{BS}}^\mathrm{H}\bold{\Theta}_{k,N}{\bold{f}_{k,N}},\\ 
\bold{h}_{k,n} = \bold{F}_{n,\text{BS}}^\mathrm{H}\bold{\Theta}_{k,n}{\bold{f}_{k,n}} + \sum\limits_{n' = n + 1}^N {\bold{F}_{n',\text{BS}}^\mathrm{H}\bold{\Theta}_{k,n}{\bold{G}_{n,n'}}} \bold{\Theta}_{k,{n'}}{\bold{f}_{k,n}},
\end{array}
\end{align}
where the channel vector between the $n$-th active STAR-RIS and the BS is denoted by $\bold{F}_{n,\text{BS}}\in\mathbb{C}^{{M} \times {N_\text{BS}}}$, while ${\bold{f}_{k,n}}\in\mathbb{C}^{{M} \times 1}$ represents the channel vector between the $k$-th user and its associated active STAR-RIS. Moreover, $\bold{G}_{n,n'}\in\mathbb{C}^{{M_n} \times {M_{n'}}}$ denotes the channel matrix from the active STAR-RIS $n$ to the active STAR-RIS $n'$. 
\subsubsection{Uplink SIC decoding} 
Without loss of generality, we assume that the users are decoded in the order of $\pi(1)$, $\pi(2)$, ..., $\pi(K)$, with their simplified effective channels expressed as follows
\begin{align}
	&{{\left|\bold{w}_{\pi({k}),n}^\mathrm{H}{\bold{h}_{\pi({k}),n}} \right|}^2}\ge {{\left|\bold{w}_{\pi(\bar{k}),n'}^\mathrm{H}{\bold{h}_{\pi(\bar{k}),n'}} \right|}^2},~
	 \forall n \ge n',~ \forall
	 {k}>\bar{k}, 
\end{align}
where $\pi({k})$ denotes the decoding order of the $k$-th user in an uplink transmission.
Moreover, it is assumed that the received beamforming vector $\bold{w}_{\pi({k}),n}\in\mathbb{C}^{{N_\text{BS}\times 1}}$ is applied at the multi-antenna BS, such that we have $\bold{w}_{\pi({k}),n}^\mathrm{H}{\bold{y}_{\text{BS}}}$. Let ${\zeta}= {{\bold{w}_{\pi({k}),n}^\mathrm{H}{\bold{h}_{\pi({k}),n}}}}$,  ${\zeta'}= {{\bold{w}_{\pi(\bar{k}),n}^\mathrm{H}{\bold{h}_{{\pi({k})},n}}}}$ and ${\zeta''} = {{\bold{w}_{\pi(j),i}^\mathrm{H}{\bold{h}_{\pi({k}),n}}}}$. Then, the signal-to-interference-plus-noise ratio (SINR) of the $[k,n]$-th user can be expressed as 
\begin{align}
\Gamma_{\pi(k),n} = \frac{{{\sum\limits_{n=1}^{{N}}}{{\beta _{\pi(k),n}}{p_{\pi(k),n}}{{\left|{\zeta} \right|}^2} + {\sum\limits_{n=1}^{N}} {(1-{\beta _{\pi(k),n}}){p_{\pi(k),n}}{{\left|{\zeta} \right|}^2}}}}}{{\sum\limits_{\bar{k}<{k}} {p_{\pi(\bar{k}),n}}{{{\left|{\zeta'} \right|}
^2}+ \sum\limits_{i=1}^{n-1}{\sum\limits_{j=1}^ {K_i}{{p}_{{\pi{(j)},i}}}{{{\left|{\zeta''} \right|}^2}}} + {{\left|\bold{w}_{\pi(k),n}^\mathrm{H} \bold{z}\right|}^2}}}}.
\end{align}
Finally, the achievable sum rate of the $[k,n]$-th user of the proposed uplink multi-active STAR-RIS-aided NOMA is given by
\begin{align}
{R_{\pi(k),n}} =  {{{{\log }_2}}} \left( {1 + \Gamma _{\pi(k),n}} \right).
\end{align}
\section{Problem Formulation and Solution}
In this section, we aim to maximize the achievable sum rate by jointly designing the power allocation matrix $\mathbf{P} = \lbrace \mathbf{p}_{\pi{(k)},n}, \forall  k,n \rbrace$, active beamforming matrix $\mathbf{W} = \lbrace \mathbf{w}_{\pi{(k)},n}, \forall  k,n \rbrace$, association indicator matrix $\boldsymbol{\beta} = \lbrace \mathbf{\beta}_{\pi(k),n}, \forall  k,n \rbrace$, and active STAR-RIS matrix $\bold{\Theta} = \lbrace \bold{\Theta}_{\pi(k),n},  \forall  k,n  \rbrace$. It is worth noting that the discrete variable  ${\beta_{\pi(k),n}}$ can be converted into a continuous variable between $0$ and $1$ by imposing the equivalent form of $\beta_{\pi(k),n} \in \lbrace 0,1\rbrace$ in $\text{C}_2$ and $\text{C}_3$.
Then, the problem $\bold{P}_{1}$ is formulated as
\begin{align} \label{ppp}
\bold{P}_{1}: \max _{\mathbf{P}, \mathbf{W}, \boldsymbol{\beta}, \bold{\Theta}} & \quad {\sum\limits_{n=1} ^{N}}{\sum\limits_{k=1}^{{K}_n}} R_{\pi(k),n}\nonumber\\
\textrm{s.t.} \quad & \text{C}_1: {R_{\pi(k),n}} \ge {R_{\min }}, \quad \forall  k,n, \nonumber \\
\quad&\text{C}_2:  0 \le \beta_{\pi(k),n} \le 1 , \quad \forall  k,n, \nonumber \\
\quad&\text{C}_3: {\sum\limits_{n=1} ^{N}}{\sum\limits_{k=1}^{{K}_n}}\left(\beta_{\pi(k),n}-\beta_{\pi(k),n}^2\right) \le 0, \nonumber \\
\quad&\text{C}_4: {\sum\limits_{n=1} ^{N}} {\beta_{\pi(k),n}}  \le 1, \quad \forall k, \nonumber\\
\quad& \text{C}_5: {{p}_{\pi(k),n}}\ge 0, \quad \forall  k,n, \nonumber\\
\quad& \text{C}_6: {\sum\limits_{n=1} ^{N}}{\sum\limits_{k=1}^{{K}_n}} {{p}_{\pi(k),n}}\le P_{\text{max}},   \nonumber\\
\quad& \text{C}_7: {{{\alpha _{m,n}^r} ,{\alpha _{m,n}^t}  \in \left[ {0,1} \right]}}, \quad \forall  m,n,   \nonumber\\
\quad& \text{C}_8: \alpha _{m,n}^r + \alpha _{m,n}^t = 1, \quad \forall  m,n,  \nonumber\\
\quad& \text{C}_9: {\theta _{m,n}^r,\theta_{m,n}^t \in \left[ {0,2\pi } \right)}, \quad \forall  m,n, \nonumber\\
\quad& \text{C}_{10}:  {{{\delta_{m,n}^r} ,{\delta_{m,n}^t}  \in \left[ {0,\delta_{\text{max}}} \right]}}, \quad \forall  m,n,  
\end{align} 
where constraint $\text{C}_1$ ensures that the minimum QoS requirements for reflection and transmission users in the $n$-th active STAR-RIS are satisfied. $\text{C}_2$ and $\text{C}_3$ indicate that the association indicator has binary value $\lbrace 0,1\rbrace$, while $\text{C}_4$ associates each user in either reflection or transmission space to at most one active STAR-RIS. Constraints $\text{C}_5$ and $\text{C}_6$ enforce such power constraint to the users. The constraints for the amplitude and phase shift coefficients of the active STAR-RISs are denoted by $\text{C}_7$, $\text{C}_8$, and $\text{C}_9$ respectively. Finally, $\text{C}_{10}$ represents the amplification gain of each element of the $n$-th active STAR-RIS.
\vspace{-0.4cm}
\section{META-DDPG ALGORITHM}
Previous studies based on convex optimization techniques utilize complex mathematical transformations to achieve the local optimum solution. However, real-time adaptability in dynamic environments has advantages which can be realized by employing DRL methods. Hence, integrating Meta-learning with the DDPG algorithm is proposed to solve problem $\bold{P}_{1}$ \cite{DDPG, Meta2, Arminmeta}. DDPG is based on an actor-critic architecture and employs a neural network where actions are determined according to a parametric strategy.  The formulated optimization problem can be modeled as a Markov decision process  $\left( {\mathcal{S},\mathcal{A},\mathcal{T},\mathcal{R},\eta } \right)$, where the state and action sets are denoted by $\mathcal{S}$ and $\mathcal{A}$, respectively. Moreover, the transition model $\mathcal{T}$ is defined as the probability of switching from the current state $s_t\in S$ at time $t$ to the next step $s'=s_{t + 1}\in S$ at time $t+1$ after taking action $a\left( {t} \right)\in \mathcal{A}$. The performance of the current state is evaluated by reward $\mathcal{R}$, while $\eta  \in \left( {0,1} \right)$ denotes a discount factor. A discount factor close to 0 emphasizes immediate rewards, whereas a discount factor close to 1 gives greater weight to long-term rewards. In general, the state serves as the input to the actor network in DDPG, which then outputs the continuous action. The critic network, in turn, takes both the action and the state as inputs to estimate the Q-value function, denoted by ${Q^Q}\left( {s,a\left| {{\omega ^Q}} \right.} \right)$. Furthermore, DDPG applies four neural networks, namely, the actor, target actor, critic, and target critic networks, with their corresponding parameters denoted by ${\omega ^\rho },{{\tilde \omega}^\rho },{\omega ^Q},{{\tilde \omega}^Q}$, respectively. In particular, the actor network generates the actions by taking the argmax and selecting actions $a = \rho \left( {s\left| {{\omega^\rho }} \right.} \right)$ in continuous action spaces, where $\rho$ represents the policy.
The combination of the actor and critic networks in DDPG defines the state-action function, which can be expressed as
\begin{align} \label{ppp2}
	Q_\rho ^Q\left( {s,a\left| {{\omega ^Q}} \right.} \right) = {\mathbb{E}_\rho }\left( {{R_t}\left| {{s_t} = s,{a_t} = a} \right.} \right),
\end{align}  
where ${R_t} = \sum\limits_{\tau  = 0}^\infty  {{\eta ^\tau }} {r^{t + \tau  + 1}}$ determines the discounted expected cumulative reward. Moreover, the following loss function, which measures the difference between the predicted value ${{Q^Q}\left( {s,a\left| {{\omega^Q}} \right.} \right)}$ and the target value according to the Bellman equation, is minimized to learn the parameters of the critic network as follows
\begin{align} \label{ppp3}
L\left( {{\omega ^Q}} \right) =  {\mathbb{E}{{\left( {X - {Q^Q}\left( {s,a\left| {{\omega ^Q}} \right.} \right)} \right)}^2}},
\end{align}
where $X = R + \left( {1 - b} \right) \eta{{\tilde Q}^Q}\left( {s',\rho '\left( {s',\rho '\left| {{\omega ^{\rho '}}} \right.} \right)\left| {{{\tilde \omega }^Q}} \right.} \right)$. Moreover, $b$ is a binary variable which represents the terminal step. Particularly, $b=1$ if the agent reaches the terminal step at each episode and $b=0$, otherwise.
\vspace{-0.3cm}
\subsection{Meta-DDPG based solution}
\begin{algorithm}[b!]
	\caption{Meta-DDPG algorithm}
	\label{algorithm1}
	\textbf{Input}: The maximum number of episodes $E_\text{max}$, the maximum number of time steps $T_\textrm{max}$, period time update $A_\text{update}$, and $0<\eta<1$. \\
	\textbf{Output}: $\left\{{ \boldsymbol{\beta}, \mathbf{P}, \mathbf{W}, \bold{\Theta}} \right\}$.\\
	Initialize the actor and critic networks as $\rho(s|\omega^\rho)$ and $Q_\rho ^Q\left( {s,a\left| {{\omega ^Q}} \right.} \right)$, respectively.\\
	Initialize the target actor and critic networks with parameters of 
	${\tilde{\omega}}^Q \leftarrow \omega^Q$ and $\tilde{\omega}^\rho \leftarrow \omega^\rho$.\\
	Initialize replay buffer $C$.
	\\
	\For{episode=1 \KwTo $E_\text{max}$}{
		initialize the environment in order to get the initial state $s$
		\\
		\For{$t=1$ \KwTo $T_\text{max}$ or $b\neq1$}{
			receiving reward $r_t$ after executing action $a_t$ in the current state. Then, the network transits from the state $s_{t}$ to the new state $s'$. The transition experience $(s_{t}, a_{t}, r_{t}, s', b)$ is stored in buffer $C$.\\  
			\For{each step of gradient descent to solve problem \eqref{ppp10}}{
				Sample a random mini batch from buffer $C$ according to the transition experience.
				\\
				$\omega^Q = \omega^Q - \text{lr}_{\text{critic}}\nabla_{\omega^Q} L(\omega^Q),$
				\\
				$\omega^\rho_{\text{old}} = \omega^\rho - \text{lr}_{\text{actor}}\nabla_{\omega^\rho} J(\omega^\rho),$
				\\
				$\omega^\rho_{\text{new}} = \omega^\rho_{\text{old}}- \text{lr}_{\text{actor}}\nabla_{\omega^\rho} \Omega_{\text{new}}(\omega^\rho,\psi ),$
				\\
				$\omega^\rho \leftarrow \omega^\rho_{\text{new}},$
				\\
				$\psi  = \psi  - \text{lr}_{\text{meta}}\nabla (\text{tanh}(J(\omega^\rho_{\text{new}})-J(\omega\theta^\rho_{\text{old}}))).$
				\\
				\If{$\text{mod}(t,A_\text{update}) = 0$}{     
					${{\tilde{\omega}}^Q}_{t+1} = (1-\upsilon){\tilde{\omega}^Q}_{t} + {\upsilon{\omega_{t}^Q}},$
					\\
					${{\tilde{\omega}}^\rho}_{t+1} =(1-\upsilon){{\tilde{\omega}}^\rho}_{t} + \upsilon{\omega_{t}^\rho}.$
				}}}}
			\end{algorithm}
In this section, the parameters of the actor vector are updated to minimize the following loss function
\begin{align} \label{ppp4}
J\left( {{\omega ^\rho }} \right) =  - \mathbb{E}\left[ {{Q^\rho }\left( {s,a} \right)\left| {s = {s_t},a = {\rho _\omega }\left( {{s_t}\left| {{{\tilde \omega}^\rho }} \right.} \right)} \right.} \right].
\end{align}
Note that the negative sign in the loss function of the actor network is applied to minimize the negative Q-value and convert the maximization problem into the minimization problem. More specifically, the actor network is encouraged to select actions that maximize the expected cumulative reward and consequently returns the higher Q-value. If $\upsilon \ll 1$, the weights of the target networks are updated periodically according to the actor and critic networks as follows 
\begin{align} \label{ppp5}
\begin{array}{l}
\tilde \omega _{t + 1}^Q \leftarrow \upsilon \omega_t^Q + (1 - \upsilon )\tilde \omega_t^Q,\\
\tilde \omega _{t + 1}^\mu  \leftarrow \upsilon \omega_t^\mu  + (1 - \upsilon )\tilde \omega _t^\mu.
\end{array}
\end{align}
The first step in solving problem $\bold{P}_{1}$ is to map it into the state space, action space, and reward framework. Following this, we will discuss this mapping along with Meta-learning.
\subsubsection{\textbf{State space}}
The system state space of the DDPG agent at time $t$, consisting of ${\bar R}$ and the channel gains at the previous time, is given by
\begin{align} \label{ppp6}
& s_t=\\ \nonumber
&\Big \lbrace{{{\bar R}^{t - 1}},{{\left\{ {\bold{f}_{k_n,n}^{t - 1}} \right\}}_{k_n \in {\mathcal{K}_n},n \in \mathcal{N}}},{{\left\{ {\bold{F}_{n,\textrm{BS}}^{t - 1}} \right\}}_{n \in \mathcal{N}}},{{\left\{ {\bold{G}_{n,n'}^{t - 1}} \right\}}_{n,n'\in \mathcal{N}}}}\Big \rbrace.
\end{align}
\subsubsection{\textbf{Action space}}
The agent takes action ${a_t} = \left\{{\boldsymbol{\beta}, \mathbf{P}, \mathbf{W}, \bold{\Theta}} \right\}$ in state $s_t$ at each time step $t$.
\subsubsection{\textbf{Reward}}
According to the optimization problem $\bold{P}_{1}$, the reward function reflects two main aspects. Firstly, it aims to optimize the total system sum rate. Secondly, it seeks to satisfy the minimum data rate criteria. Let ${\vartheta _{\pi(k),n}}$ denote the penalty term designed to penalize any user in the proposed system model whose current transmission rate does not meet the minimum transmission rate requirements as follows
\begin{align} \label{ppp8}
{\vartheta _{\pi(k),n}} = \left\{ \begin{array}{l}
0, \quad \quad \quad \quad {R_{\pi(k),n}} \ge {R_{\min }},\\
- {R_{\pi(k),n}}, \quad  0 \le {R_{\pi(k),n}} \le {R_{\min }},\\
-{c_3}, \quad \quad \quad \quad {R_{\pi(k),n}} = 0.
\end{array} \right.
\end{align} 
Then, the reward function can be formulated as follows
\begin{align} \label{ppp7}
{\bar R} = {c_1}{\sum\limits_{k=1}^{{K}_n}}{\sum\limits_{n=1} ^{N}} R_{\pi(k),n} + {c_2}{\sum\limits_{k=1}^{{K}_n}}{\sum\limits_{n=1} ^{N}}{{\vartheta _{\pi{(k)},n}}},
\end{align} 
where ${c_1}$, ${c_2}$, and ${c_3}$ are positive constant coefficients and are chosen according to the importance of different parts of reward function.
Finally, the immediate reward which forces the proposed learning model to satisfy the constraints can be described as
\begin{align} \label{ppp9}
{r^t} = \left\{ \begin{array}{l}
\bar R, \quad \quad  \text{if constraints are satisfied,}\\
0, \quad \quad \text{otherwise}.
\end{array} \right.
\end{align} 
The DDPG algorithm has limitations in quickly adjusting to a new and dynamic environment. To address this concern, we propose the integration of DDPG with Meta-learning. In the following, the Meta-DDPG algorithm is explained in detail. First of all, a bi-level optimization problem is defined as follows
\begin{align}
\label{ppp10}
\begin{aligned}
\psi  &= \arg \min_\psi \Omega_{\text{meta}}(\omega^{*\rho}) \\
\text{s.t.} \quad \omega^{*\rho} &= \arg \min_{\omega^\rho} \left(J(\omega^\rho) + \Omega_{\text{new}}(\omega^\rho, \psi )\right).
\end{aligned}
\end{align}
At the outer level, the Meta knowledge $\psi \in [0,1]$ is obtained through the minimization of the Meta loss function, ${\Omega _{\text{meta}}}\left( {{\omega ^{ * \rho }}} \right) = \tanh \left( {J\left( {\omega _{\text{new}}^\rho } \right) - J\left( {\omega _{\text{old}}^\rho } \right)} \right)$ \cite{Meta2}. Moreover, at the inner level, a new loss function denoted by ${J\left( {{\omega ^\rho }} \right) + {\Omega _{new}}\left( {{\omega ^\rho },\psi  } \right)}$ is minimized by the actor parameters, where ${\Omega _{\text{new}}}\left( {{\omega ^\rho },\psi  } \right) = \mathbb{E}\left[ {\psi \left( {\log \left( {1 + {e^{{\rho _\omega }\left( {s\left| {{\omega ^\rho }} \right.} \right)}}} \right)} \right)} \right]$. The actor parameters that are updated by conventional DDPG and Meta-DDPG are defined as $\omega _{\text{old}}^\rho  = {\omega ^\rho } - {\text{lr}_{\text{actor}}}{\nabla_{{\omega^\rho }}}J\left( {{\omega^\rho }} \right)$ and $\omega _{\text{new}}^\rho  = \omega_{\text{old}}^\rho  - {\text{lr}_{\text{actor}}}{\nabla_{{\omega ^\rho }}}{\Omega _{\text{new}}}\left( {{\omega ^\rho },\psi} \right)$, respectively.
\subsection{Time-Complexity Analysis of the Meta-DDPG Algorithm}
In this section, the time complexity of the proposed Meta-DDPG algorithm is analyzed by adapting the floating point operations per second (FLOPS) for the hidden layers of the actor, critic, and actor-critic networks \cite{Wang}. It is assumed that the actor and critic networks contain $\mathcal{I}$ and $\mathcal{J}$ connected layers, respectively, while the Meta-critic network contains $\mathcal{L}$ connected layers. By taking the activation layer with associated parameter $\upsilon$ into account, the time complexity of the actor network can be computed as $2\sum\nolimits_{i = 0}^\mathcal{I} {\left( {\left( {2{\zeta _{a,i}} - 1} \right){\zeta _{a,i + 1}} + \upsilon {\zeta _{a,i + 1}}} \right)}$. Similarly, the time-complexity of the critic and Meta-critic networks are computed, in which the overall time-complexity of the proposed Meta-DDPG algorithm can be expressed as
\begin{align}
\mathcal{O}\left( {\sum\limits_{i = 0}^{\mathcal{I} - 1} {{\zeta _{a,i}}{\zeta _{a,i + 1}} + \sum\limits_{j = 0}^{\mathcal{J} - 1} {{\zeta _{c,j}}{\zeta _{c,j + 1}} + \frac{1}{2}\sum\limits_{l = 0}^{\mathcal{L} - 1} {{\zeta _{m - c,l}}{\zeta _{m - c,l + 1}}} } } } \right),
\end{align} 
while the time-complexity of DDPG algorithm is given by
\begin{align}
\mathcal{O}\left( {\sum\limits_{i = 0}^{\mathcal{I} - 1} {{\zeta _{a,i}}{\zeta _{a,i + 1}} + \sum\limits_{j = 0}^{\mathcal{J} - 1} {{\zeta _{c,j}}{\zeta _{c,j + 1}}}}} \right).
\end{align} 
A trade-off between complexity and adaptation is considered.
While Meta-DDPG algorithm increases complexity by $2.46\%$
compared to DDPG, it offers significantly better adaptability.
\section{Simulation Results}
\begin{figure*}[t]
	\centering
	\begin{minipage}[b]{0.24\textwidth}
		\centering
		\includegraphics[width=\linewidth]{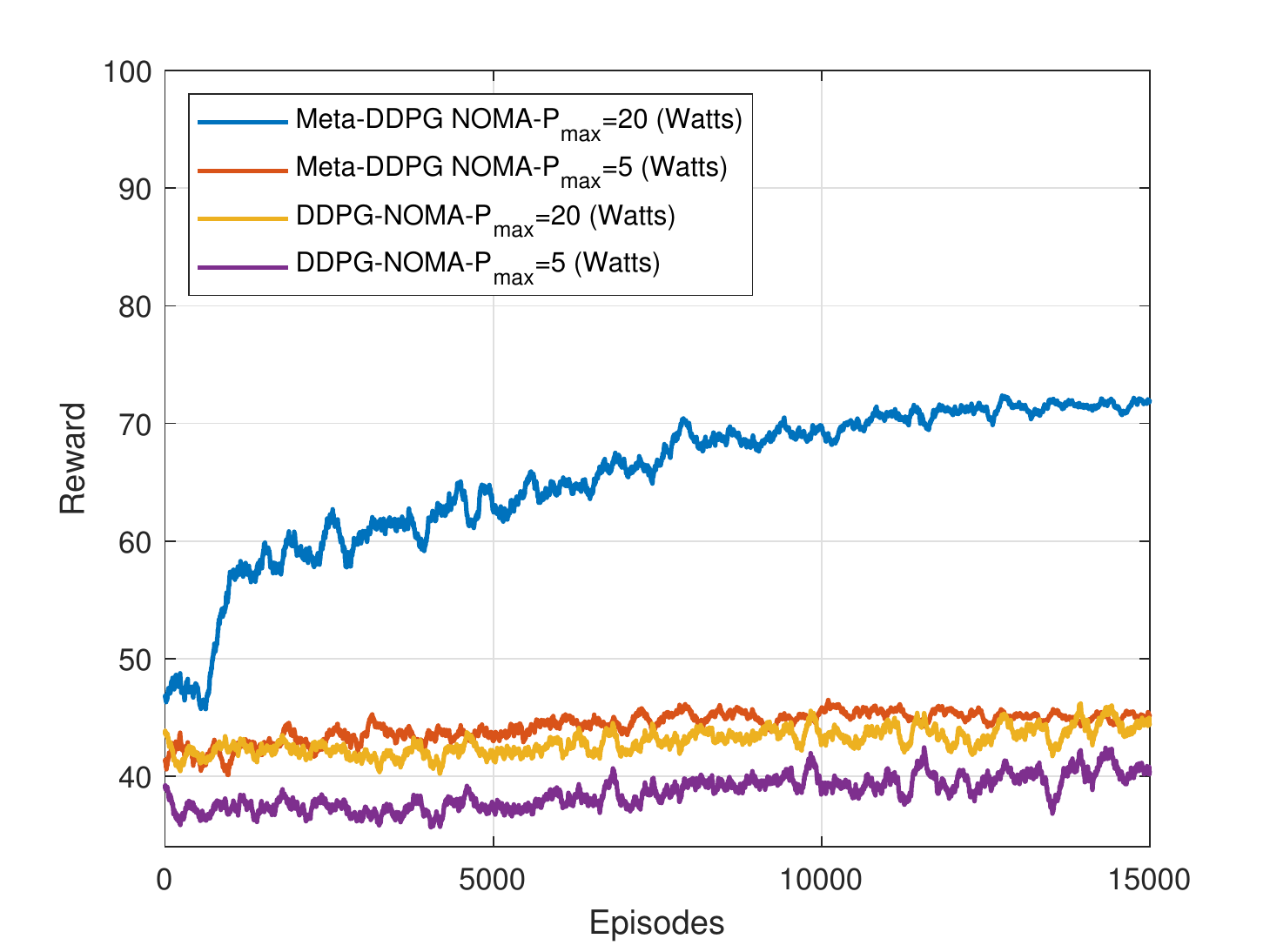}
		\caption{\small The reward in different episodes.}
		\label{fig1}
	\end{minipage}%
	\hfill
	\begin{minipage}[b]{0.24\textwidth}
		\centering
		\includegraphics[width=\linewidth]{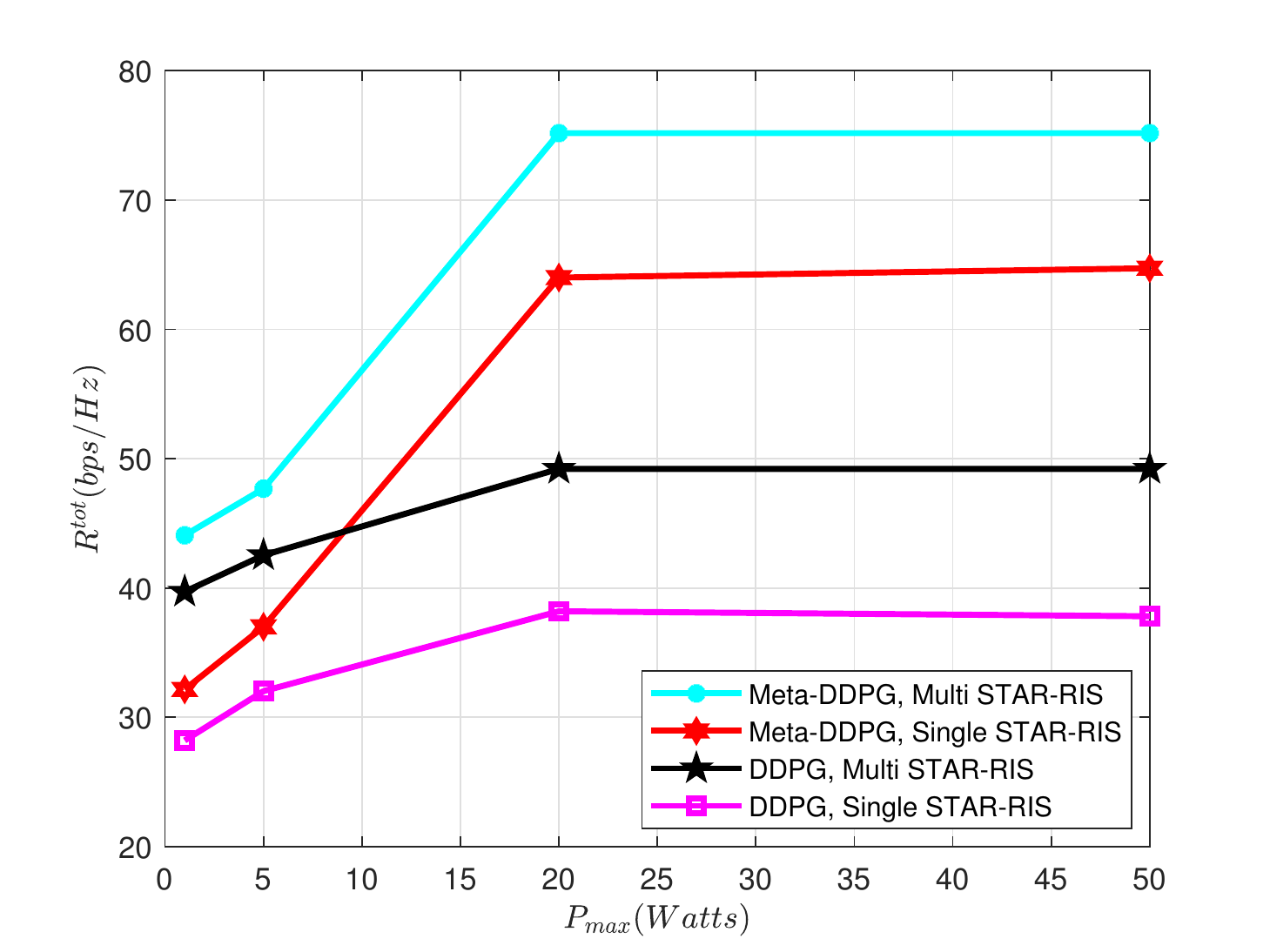}
		\caption{\small The system's total rate in different episodes.}
		\label{fig2}
	\end{minipage}%
	\hfill
	\begin{minipage}[b]{0.24\textwidth}
		\centering
		\includegraphics[width=\linewidth]{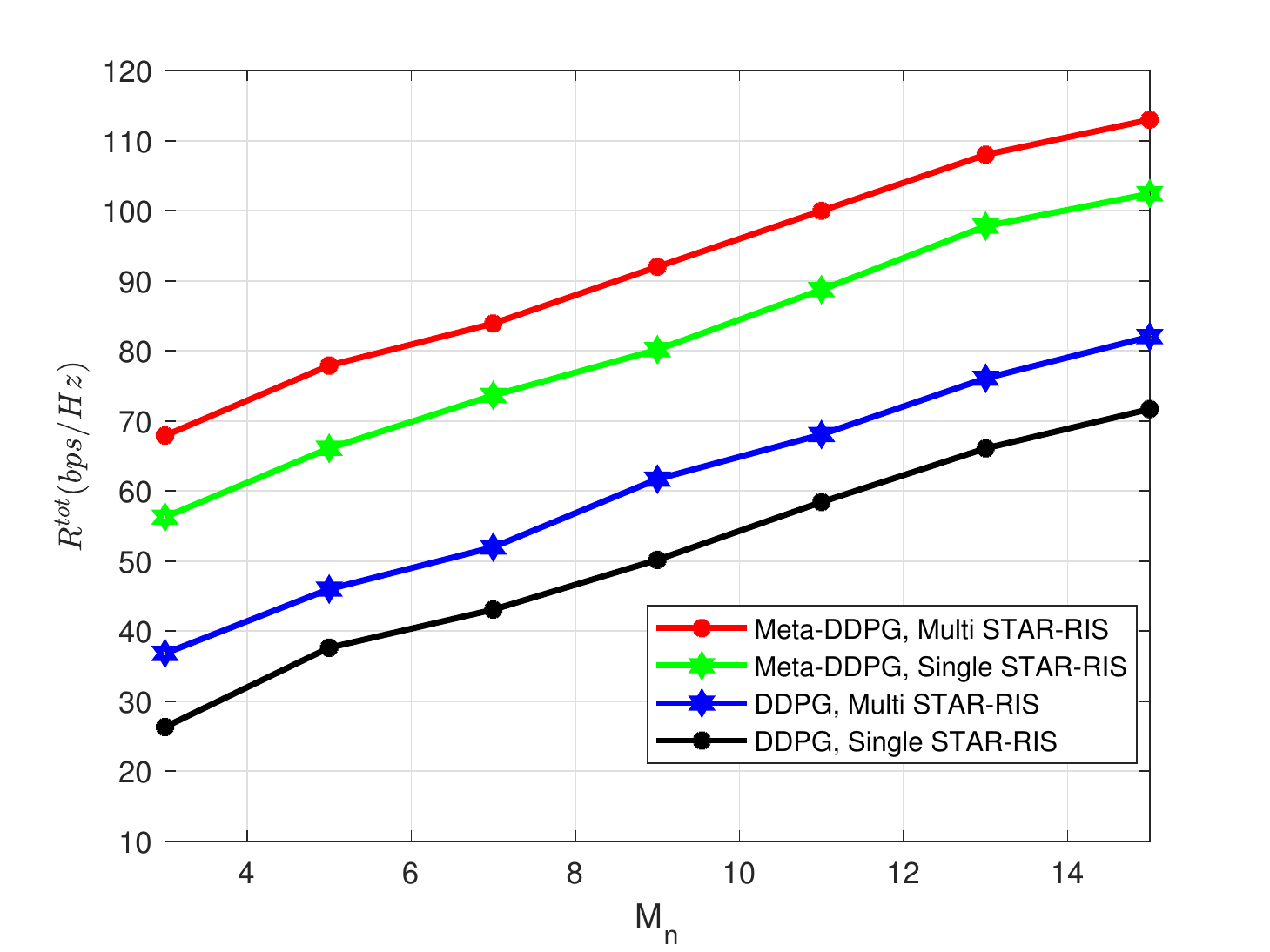}
		\caption{\small The system's total rate versus $P_{\text{max}}$.}
		\label{fig3}
	\end{minipage}%
	\hfill
	\begin{minipage}[b]{0.24\textwidth}
		\centering
		\includegraphics[width=\linewidth]{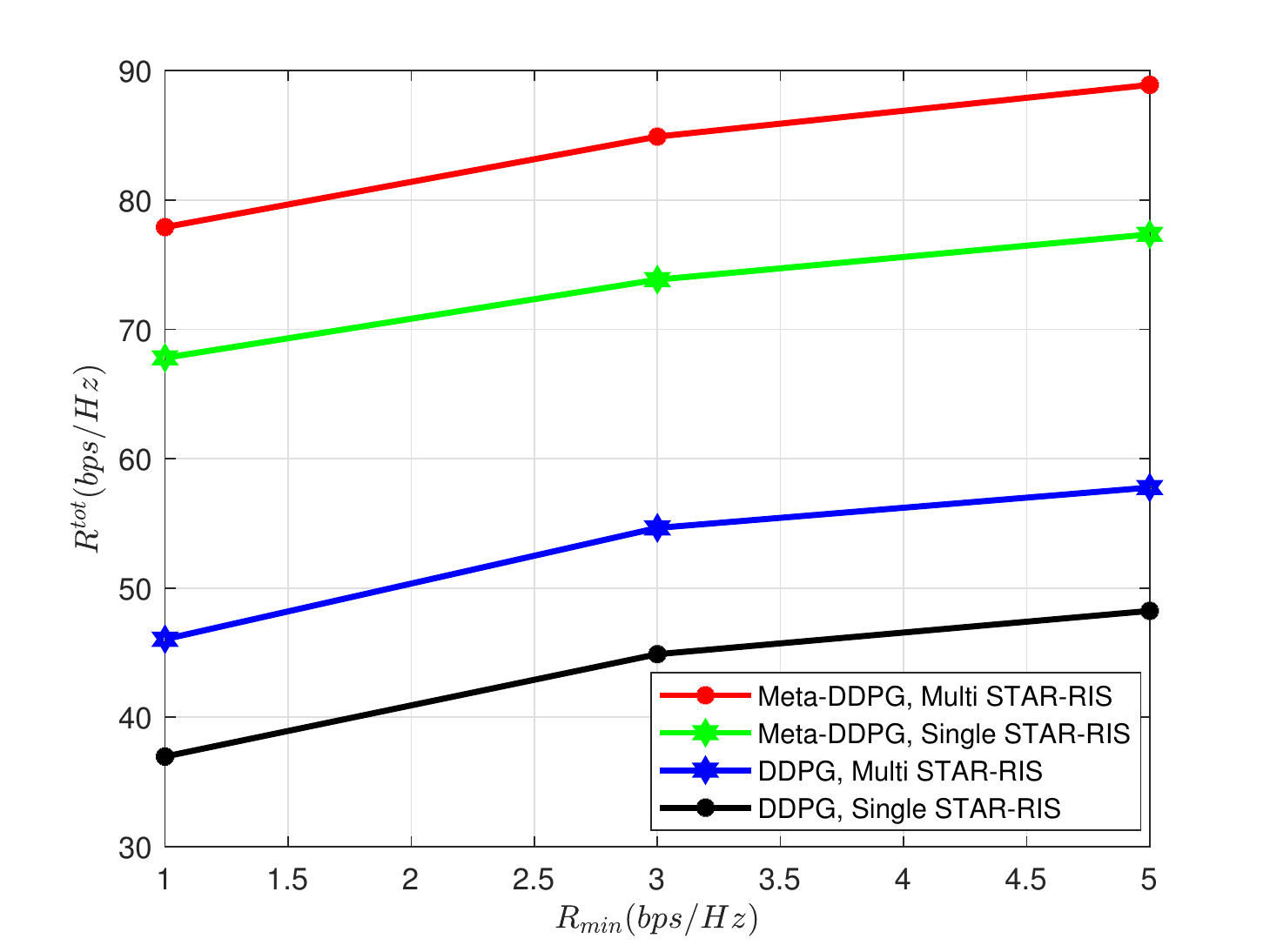}
		\caption{\small{The system's total rate versus the number of elements.}}
		\label{fig4}
	\end{minipage}
\end{figure*}

In this section, numerical simulations are conducted to assess the performance of the proposed Meta-DDPG algorithm. The simulation parameters and hyperparameters for Meta-DDPG in Rayleigh fading channels are detailed in Table \ref{tab:my_label}, unless specified otherwise. \\
Fig. \ref{fig1} illustrates the convergence and performance comparison between the Meta-DDPG and DDPG algorithms, highlighting the superiority of multi-active STAR-RIS over multi STAR-RIS across varying numbers of episodes. Specifically, the results show that multi-active STAR-RIS  outperforms multi STAR-RIS due to amplification gains, while the Meta-DDPG algorithm consistently outperforms the DDPG algorithm with $19\%$ improvement.\\
Fig. \ref{fig2} compares the performance of the proposed system model with secondary reflections against single reflections, considering different values of $P_{\text{max}}$ across various episodes. The results demonstrate that the multi-active STAR-RIS with second-order reflections outperforms the single-reflection with $74.1\%$ enhancement in the total data rate.
\\
Fig. \ref{fig3} provides a visualization of the relationship between the total data rate and different values of $P_{\text{max}}$ for both multi-active STAR-RIS and single-active STAR-RIS scenarios. In this comparison, it is apparent that an increase in the level of $P_{\text{max}}$ leads to an improvement in the total data rate, while employing the multi-active STAR-RIS scenario results in higher total data rate compared to the single-active STAR-RIS scenario. Moreover, a comparison between the Meta-DDPG algorithm and the convex approach for multi-active STAR-RIS is presented. The results show that Meta-DDPG provides a strong approximation to the mathematical convex method. Specifically, the mathematical complexity is denoted as $\mathcal{O}(J_\text{AO}( J_1\mathcal{O}_1+J_2\mathcal{O}_2+J_3\mathcal{O}_3))$, where $\mathcal{O}_1\left( {\sqrt u \log \left( {1/\zeta } \right)\left( {x{u^3} + {x^2}{u^2} + {x^3}} \right)} \right)$ and ${\mathcal{O}_2}(\max (N{M_n},{\log _2}(1/{\varepsilon_1}){(2K)^4}\sqrt {N{M_n}}))$, with $x=12K$, $u=N_\text{BS}^2$, and $\varepsilon_1$ being the solution accuracy. We define ${\mathcal{O}_3}(\left[ {{{\log }_2}(\left( {KN + 1} \right)/{t^0}{\varepsilon_2})} \right]/\left[ {{{\log }_2}\varsigma + KN} \right])$, with ${t^0}$, ${\varepsilon_2}$, and $\varsigma$ being the initial point, the accuracy, and the gradient of step size, respectively. Furthermore, $J_\text{AO}$, $J_1$, $J_2$, and $J_3$ represent the corresponding number of iterations. As can be seen, due to iterations, the total complexity of the convex benchmark is very high, and in this regard, the proposed solution outperforms convex optimization.\\
Finally, Fig. \ref{fig4} depicts the impact of the number of active STAR-RIS elements on the total data rate for different numbers of users. From this graphical representation, it becomes evident that as the number of active STAR-RIS elements increases, there is an improvement in the total data rate. Furthermore, it is evident that multi-active STAR-RIS scenario has superior performance compared to the multi STAR-RIS scenario.
\begin{table}[t!] 
	\vspace{-1 em}
	\centering
	\caption{Simulation Parameters}
	\label{tab:my_label}
	\vspace*{5pt}
	\begin{tabular}{ l|l||l|l }
		\textbf{Parameter} & \textbf{Value} & \textbf{Parameter} & \textbf{Value} \\
		\hline
		\hline		
		\hline
		$P_\text{max}$   & $~\text{20 Watts}  $ &
		$E_\text{max}$ &  5000\\ 
		\hline $ K $ & 16 &
		\text{Discount factor}& 0.99 \\ 
		\hline $ R_{\text{min}} $ & $ 1~\text{bps/Hz} $ &
		\text{Noise power}& -174 dBm/Hz \\
		\hline $ \mid B \mid $ & $ 10^4 $  &\text{Batch size}& 100  \\
		\hline $N_\text{BS}$ & 4 &
		$N$ & 4 \\
		\hline $\delta_\text{max}$ & 25 \text{dB}&
	\end{tabular}
	\label{tab2}
\end{table}
\section{Conclusion} 
In this letter, the resource allocation scheme in a multi-active STAR-RIS-aided system is investigated, where the users apply NOMA to assist the efficient transmission. More specifically, the sum rate maximization problem is formulated by jointly optimizing the active beamforming, power allocation, transmission and reflection beamforming at the active STAR-RIS, and user- active STAR-RIS association indicator. Simulation results demonstrated the effectiveness of the Meta-DDPG learning model, while the higher data rate is achieved by considering second-order reflections among multiple active STAR-RISs. Moreover, the superiority of the proposed multi-active STAR-RIS over multi STAR-RIS is verified.
{\color{white}\subsection{}}
\vspace{-0.5cm}
\bibliographystyle{IEEEtran}
{\footnotesize
\bibliography{IEEEabrv,reference}
\end{document}